\documentclass[12pt]{article}
\title{Criteria for Exact Solubility of Relativistic Field Theories by 
Scattering Transform}
\author{Gautam Bhattacharya\footnote{
e-mail address: gautam@theory.saha.ernet.in}\\
Theory Group\\
Saha Institute of Nuclear Physics\\
1/AF Bidhannagar, Calcutta 700 064\\
INDIA}
\date{}
\begin{document}
\maketitle
\begin{abstract}
Scattering transform is a well known powerful tool for quantisation of field 
theories in (1+1) dimensions. Conventionally only those models whose classical
counterparts admit a Lax pair (origin of which is always mysterious) have been 
quantised in this way. In relativistic quantum field theories we show that
the scattering transforms can be constructed ab initio from its invariance
under Lorentz 
transformation (both proper and improper), irreducible transformation nature
of scalar and Dirac fields, the existence of a momentum scale associated
with asymptotic nature of the scattering transform and the closure of short
distance operator product algebra. For single fields it turns out that 
theories quantisable by scattering transforms are restricted to sine-Gordon
type for spin-$0$ and Massive Thirring type for spin-$\frac{1}{2}$ if the
target space of the scattering transform matrix is assumed to be parity
invariant. There are interesting unexplored extensions if the target space 
is given chirality. 

\end{abstract}

\thispagestyle{empty}
\vspace{36pt}
\noindent PACS No. : 11.10.Kk, 11.10.Lm, 11.25.Sq

\vspace {36pt}
\noindent Keywords : Scattering transform, Lorentz covariance, Operator 
product algebra, Braiding matrix, Master equation, Schwinger term

\newpage
Just as free field theories can be exactly solved by showing that the 
Hamiltonian, expressed in terms of the components of the Fourier transforms of
the fields and their canonical conjugates, is cleanly diagonalisable -- quite
a few nonlinear field theories are also known to be integrable in an equivalent 
sense
that a certain scattering transform give rise to a set of new variables in
terms of which the Hamiltonian is exactly diagonalisable\cite{Faddeev}. These 
transforms
are constructed from a constant time path ordered integral (Wilson line 
\cite{Wilson1}) of a local Lie algebra valued
object called the Lax operator (or the scattering operator since, for $x\rightarrow
\pm\infty$, it takes a value of a constant momentum $k$ and hence the 
integral itself has asymptotic plain wave structure). Expressed mathematically,
the scattering matrix ${\cal T}(x,y;k)$ is given by
\begin{equation}
{\cal T}(x,y;k)={\cal P}\exp\left[i\int^x_y{L(\xi,k)d\xi}\right]\label{monodromy}
\end{equation}
\begin{equation}
L(x,k)=\sum_i^N{t^i(k)j_i(x)}\rightarrow T_dk {\rm ~~as~} x\rightarrow\pm\infty
\label{Lax}
\end{equation}
Here $t^i(k)$ are the generators of a graded  algebra in the form
\begin{equation}
t^i(k)=\sum_a{f_a^i(k)t^a}\label{grade}
\end{equation}
where $t^a$'s are generators of a simple Lie algebra
(normally taken in its 
fundamental representation for simplicity in algebraic manipulation) and 
$f_a^i(k)$ are appropriately chosen functions of the momentum $k$.
$T_d$ is 
in the Cartan sub-algebra (hence taken diagonal). The dynamical information are
all in the $N$ number of local fields $j_i$'s characterised by their non-trivial equal
time commutation algebra

It is known \cite{Faddeev,Kulish} that for some appropriate choices of $L(x,k)$ the 
scattering marix
would satisfy the following braid algebra 
\begin{equation}
{\cal R}(k,q)[{\cal T}(x,y;k)\otimes I][I\otimes{\cal T}(x,y;q)]
=[I\otimes{\cal T}(x,y;q)][{\cal T}(x,y;k)\otimes I]{\cal R}(k,q)\label{braid}
\end{equation}
Here ${\cal R}$, called the braiding matrix that acts on the product 
representation space of the
two scattering transforms, depends on the spectral variables $k$ and $q$ only.
This automatically lead to the involution relation
\begin{equation}
\left[{\rm Trace}({\cal T}(k)),{\rm Trace}({\cal T}(q))\right]=0
\label{involution}\end{equation}
where ${\cal T}(k)$'s are the actual scattering matrics in the sense that
one has taken
$x\rightarrow\infty$ and $y\rightarrow-\infty$ with the asymptotic oscillating 
plane wave components factored out. This involution algebra leads to the 
existence of infinite number of mutually commuting objects which are the
coefficients of different powers of $k$ after Taylor expanding 
${\rm Trace}({\cal T}(k))$ in (inverse) powers of $k$. One of these 
coefficients when evaluated as an integral in terms of the basic field 
variables (usually the coefficient of $k^{-2}$ for non-relativistic theories)
would look like a Hamiltonian
of a nonlinear theory. This means that such a Hamiltonian system has infinite 
number of constants of motion, all mutually commuting. In that sense these 
quantum field theories are integrable. The explicit form of
${\cal R}$ also enables one to study the evolution pattern for the non-diagonal
operators of the scattering matrix ${\cal T}$ by studying their commutation
algebra with the diagonal elements. These non-diagonal operators
are the counterparts of the Fourier transforms and hence are appropriately 
called the {\em scattering transforms}.  All the eigenvalues any of 
the conserved quantities including the Hamiltonian can subsequently be found
from the commutation algebra between the constants of motion and the scattering
transforms \cite{Faddeev}.

Exactly what forms of $L(x,k)$ would finally lead to the involution relation
like Eq.(\ref{involution}) has always remained a big mystery.  Traditionally
one picks up those classical nonlinear models which admit Lax pairs \cite{AKNS}
 and choose
the one corresponding to the space variation as the $L$ operator of 
Eq.(\ref{Lax}). What we establish in this paper is that in the relativistic 
theories in 1+1 dimensions the choices for such $L$ cannot be too
many. For this one need not start with any such classically integrable system
but use the properties of local quantum fields and arrive at a consistent form
of $L$ that would satisfy Eq.(\ref{braid} and hence the integrability (the
classical limit of Eq.(\ref{involution})). We show now how 
the following conditions put such severe restriction on the form of $L$.

\vspace{24pt}
\noindent {\em Relativistic covariance:}

It is clear from Eq.(\ref{monodromy}) that the scattering matrix is
a connection term - something like a Wilson line in a non-Abelian gauge theory
and for the truly infinite limit it has to be a Lorentz invariant quantity. 
Moreover, to clearly distinguish between momentum and Hamiltonian from their 
Lorentz covariance we have to make ${\cal T}(x,y;k_1)$ parity and time reversal 
invariant too. {\em This means $L(x,k_1)$ has to be chosen in such a way that it
would transform like the space component of a true vector}. Here we are assuming
that the target space (the representation space on which ${\cal T}$ acts) is
parity invariant. In the later part of the paper we will relax this condition
to explore the possibilities of existence of other soluble models.

The next thing that one has to remember that each fundamental local field must
transform irreducibly under Lorentz transformation (LT) that itself acts 
irreducibly on the light cone variables,
\begin{equation} 
x^0\pm x^1=x_\pm\rightarrow e^{\pm\theta}x_\pm
\end{equation}
where $\theta$ is the boost parameter. Scalar fields are invariant under LT
whereas Dirac fields transform as 
\begin{equation}
\psi_{1,2}\rightarrow e^{\pm\theta/2}\psi_{1,2}
\label{spinor}
\end{equation}
Finally one has to think about the LT of the spectral parameter that enters into
the picture from the asymptotic behaviour where the fields are supposed to 
vanish in the matrix element sense. Instead of taking $k_1$, we will take the
irreducible objects $k_0\pm k_1=k_\pm$ with $k_+k_-$ a Lorentz invariant 
constant and that we identify with a given mass scale of the theory. Hence
the appropriate dimensionless spectral parameter is taken to be $\lambda$ such
that
\begin{equation}
k_+=m\lambda ~~{\rm and}~~k_-=\frac{m}{\lambda}
\end{equation}
One important consequence of Lorentz covariance is that the grading 
functions $f^i_a (\lambda)$ occurring in $L$ must be a monomial in $\lambda$
with the power 
determined by the nature of LT of the Lorentz irreducible component $j_i(x)$
of Eq.(\ref{Lax}).
For example, if $j_i(x)$ has `spin' $n_i$ i.e., it transforms 
irreducibly under LT as
\begin{equation}
j_i\rightarrow e^{n_i\theta}j_i
\end{equation}
the associated grading function must transform irreducibly as 
$$f^i_a(\lambda)\rightarrow e^{(-n_i\pm 1)\theta}f^i_a(\Lambda^{-1}\lambda)$$
implying
\begin{equation}
f^i_a(\lambda)=c^i_a\lambda^{-n_i \pm 1}
\end{equation}
Since this is multiplied by the generators $t^a$ of a Lie algebra, the constants
$c^i_a$'s just cause a linear combination of the old basis $t^a$ to a new
basis $T^i$ and in Lax operator these are just graded by a single power of the
spectral parameter - a considerable reduction from the original general form
suggested in Eq.(\ref{grade})!
The changed basis will be complete if the number of fields ($N$) is exactly
equal to
the dimension of the Lie algebra. However, this matching is not necessary. We
may have a under-complete basis (as in Toda field theory) or an over-complete
basis (as in massive Thirring model to be discussed in this paper itself).

\vspace{24pt}
\noindent {\em Parity Invariance:}

If one insists on parity invariance, i.e. if one demands that under parity
$L(x,\lambda) \rightarrow - L(x,\lambda)$, one can remove the $\pm 1$ ambiguity
in the grading. $L$, like the space component of momentum, does not transform
irreducibly under LT. It must have one part that will transform like $\lambda$
and another that transforms like $1/\lambda$.  In other words
\begin{equation}
L(x,\lambda)=L_+(x,\lambda)-L_-(x,\lambda)\label{11}
\end{equation}
with $L_\pm(x,\lambda)\rightarrow e^{\pm\theta}L_\pm(\Lambda^{-1}x,\lambda)$
and
 $L_+\longleftrightarrow L_-$ under parity. In such a situation
 we can grade separately the 
 generators of $L_\pm$
 by the functions $f^i_{a,\pm}(\lambda)$ and consequently, using the LT
 restriction --
\begin{equation}
\begin{array}{rcl}
f^i_{a,+}(\lambda)&=&c^i_a\lambda^{-n^+_i + 1}\\
f^i_{a,-}(\lambda)&=&c^i_a\lambda^{-n^-_i - 1}\\
\end{array}
\end{equation}
where $n^\pm_i$ are the `spin' indices of the parity conjugate fields $j^\pm_i$,
$j^+_i\stackrel{\rm parity}{\longleftrightarrow} j^-_i$ and
\begin{equation}
L_\pm(x,\lambda)=f^i_{a,\pm}(\lambda)t^aj^\pm_i(x)
\end{equation}

\vspace{24pt}
\noindent{\em Causality:}

One main reason for taking a constant time path for the scattering transform
is to ensure that all local fields are causally separated. This makes the 
operator ${\cal T}(x,y;\lambda)$ well defined in quantum theory. The quantum
dynamics enter the picture when we take exterior product of one
${\cal T}(x,y;\lambda)$ with another to check the existence of braiding
relation. This is where we will come across products of fields in the equal
time short distance limit.  To have a non-trivial braiding (i.e. getting a
${\cal R}(\lambda,\mu)$ different from identity matrix and depending on
the spectral parameters), we must encounter
products that would have singularities in the equal time short distance limit.
This can happen only when $L(x,\lambda)$ contains canonically conjugate
fields, their polynomials and even entire functions. One should note here that
we are translating the equal time commutator, generally used in non-relativistic
integrable systems, into
a more symmetric looking operator product expansion with singularities.

\vspace{24pt}
\noindent{\em Closure of  Operator Product Algebra:}

The choice of fields to be included in $L(x,\lambda)$ must satisfy another 
important condition, namely the {\em closure} under operator product expansion
(OPE)\cite{Wilson}
\begin{equation}
T(j_i(x)j_j(y))=\sum_{n\ge 1}{{{^{(n)} O_{ij}(x)}\over{(x-y-i0^+)^n}}}
+\ \ {\rm regular\ terms} 
\label{ope}
\end{equation} 
The closure implies that all the local operators $^{(n)}O_{ij}(x)$
(including identity) associated
with the pole singularities of any pair of fields occurring in $L(x,\lambda)$ 
must 
themselves occur in $L(x,\lambda)$. To prove this point, 
consider the products of 
two infinitesimal Wilson lines ${\cal T}_\epsilon(x;\lambda)$ and 
${\cal T}_\epsilon(x;\mu)$, where 
\begin{equation}
{\cal T}_\epsilon(x;\lambda)\equiv {\cal T}(x+\epsilon/2,x-\epsilon/2;k)
=1+i\int^{x+\frac{\epsilon}{2}}_{x-\frac{\epsilon}{2}}{t^i(\lambda)
j_i(\xi)d\xi} + {\cal O}(\epsilon^2)
\end{equation}
If there were no singularities in the OPE, this product, to order $\epsilon$,
will have exactly the same look as the classical product, namely
$$
\begin{array}{rcl}
[{\cal T}_\epsilon(x;\lambda) \otimes 1]
[1 \otimes {\cal T}_\epsilon(x;\mu)]& = &
1\otimes 1 + i[(t^i(\lambda)\otimes 1)+(1\otimes t^i(\mu))]
{\displaystyle\int^{x+\frac{\epsilon}{2}}_{x-\frac{\epsilon}{2}}
{j_i(\xi)d\xi}}\\
&=&[1 \otimes {\cal T}_\epsilon(x;\mu)][{\cal T}_\epsilon(x;\lambda)
\otimes 1]
\end{array}
$$
implying trivial braiding (${\cal R}=1$).
When there are singularities in the OPE, there will be more terms to order
$\epsilon$ \cite{gbsg1}. For example, 
$$(t^i(\lambda)\otimes t^j(\mu))\int^{x+\frac{\epsilon}{2}}_{x-\frac{\epsilon}
{2}}
j_i(\xi)d\xi
\int^{x+\frac{\epsilon}{2}}_{x-\frac{\epsilon}{2}}
j_j(\eta)d\eta$$
which was classically of order $\epsilon^2$ and hence ignored, will now be
of order $\epsilon$ if $j_i(\xi)j_i(\eta)$ has a first order pole singularity.
If the local field associated with this singularity is $^{(1)}O_{ij}$ 
then, to order $\epsilon$, one will have

\begin{equation}
\begin{array}{lcl}
[{\cal T}_\epsilon(x;\lambda) \otimes 1]
[1 \otimes {\cal T}_\epsilon(x;\mu)]
&=&
1\otimes 1 + i
{\displaystyle\int^{x+\frac{\epsilon}{2}}_{x-\frac{\epsilon}{2}}
{i[(t^i(\lambda)\otimes 1)+(1\otimes t^i(\mu))]
j_i(\xi)d\xi}}\\
&&+c
{\displaystyle\int^{x+\frac{\epsilon}{2}}_{x-\frac{\epsilon}{2}}{
(t^i(\lambda)\otimes t^j(\mu))^{(1)}O_{ij}(\xi)d\xi}},\\
{\rm whereas}&&\\

[ 1 \otimes {\cal T}_\epsilon(x;\mu)]
[{\cal T}_\epsilon(x;\lambda) \otimes 1]
&=&
1\otimes 1 + i
{\displaystyle\int^{x+\frac{\epsilon}{2}}_{x-\frac{\epsilon}{2}}
{i[(t^i(\lambda)\otimes 1)+(1\otimes t^i(\mu))]
j_i(\xi)d\xi}}\\
&&+c
{\displaystyle\int^{x+\frac{\epsilon}{2}}_{x-\frac{\epsilon}{2}}{
(t^i(\mu)\otimes t^j(\lambda))^{(1)}O_{ij}(\xi)d\xi}}\\
\end{array}\label{15}
\end{equation}

Clearly ${\cal R}$ is not proportional to identity any more. However if 
${\cal R}$ exists it must satisfy a relation
$$
\begin{array}{l}
[{\cal R}, (t^i(\lambda)\otimes 1)j_i(x)+(1\otimes t^i(\mu))j_i(x)]\\
-ic\left({\cal R}t^i(\lambda)\otimes t^j(\mu)^{(n)}O_{ij}(x) - t^i(\mu)\otimes t^j(\lambda)
^{(1)}O_{ij}(x){\cal R}\right)
=0
\end{array}
$$
Now if $^{(1)}O_{ij}(x)$ are linearly independent of $j_i$'s one will have two sets
of algebraic equations for ${\cal R}$,
\begin{equation}
\begin{array}{l}
[{\cal R}, (t^i(\lambda)\otimes 1)+(1\otimes t^i(\mu))]=0\\
{\cal R}t^i(\lambda)\otimes t^j(\mu) - t^i(\mu)\otimes t^j(\lambda)
{\cal R}=0\\
\end{array}
\end{equation}
The second relation is inconsistent. In the classical limit when ${\cal R}$ is
close to identity, to the lowest order in Plank constant one would
get
$$t^i(\lambda)\otimes t^j(\mu) - t^i(\mu)\otimes t^j(\lambda)=0$$
a relation that can only be true when $\lambda=\mu$ since the power
of the monomials are necessarily different.

When the leading singularities of OPE  correspond to poles higher than
first order the 
calculation of the quantum correction is not as straightforward as has been 
done in Eq.(\ref{15}). We will
discuss it in the latter part of the article when we will show that by
appropriate gauge transformation of the Wilson line, the singularities of
the OPE among the relevant operators in the gauge transfomed $L$ can be 
always reduced to simple poles. 
Consequently the same argument as discussed above can be invoked to prove 
that the local fields $j_i$s occurring in $L(x,\lambda)$ must satisfy a 
closed OPE, namely,
\begin{equation}
T(j_i(x)j_j(y))=\sum_n{\displaystyle{\frac{^{(n)}F^k_{ij}\ j_k(x)}{(x-y-i0^+)^n}}}
+~~{\rm
regular~terms}
\label{closure}
\end{equation}

\vspace{24pt}
\noindent {\em Construction of Lax Operators:}

Equipped with the necessary requirements of Lorentz covariance, causality, 
and closure of local fields in OPE, we now proceed to actual construction
of $L$ operators with `spin' $0$ (scalar or pseudo-scalar) and `spin' $\frac
{1}{2}$ (Dirac) fields separately. For the sake of simplicity we will not
include any flavour index, though such extensions are incorporable with 
corresponding enlargement of the Lie algebra.

Consequent to the discussion in the last part of the previous section, we will 
also confine ourselves, for the time being, to the construction of those 
$L$ for which OPE of the local fields do not have singularities higher than 
first order poles.

\vspace{12pt}
\noindent {\em Spin-$0^+$:}

For true scalar fields we encounter a no go situation. Recall that $L$ must
contain both $\phi$ and $\dot\phi$ so that the equal time limit of the OPE of 
$L$'s would be nontrivial (i.e. having singularities). $\dot\phi$ which
is the canonical conjugate of $\phi$ transforms as the time component of a 
vector and cannot occur in $L$ in a simple way. One can try a construed way
of introducing a term like $k_1(k.\partial\phi)$ or $k_1(\partial \phi)^2/m^2$ 
but that will violate the closure property.
We could have added a term like $\epsilon_{10}\partial^0\phi$ in 
$L$ but that will destroy the parity invariance.

\vspace{12pt}
\noindent {\em Spin-$0^-$ (Pseudo-scalar)}

For a pseudo-scalar field $\phi$, all even functions will be parity invariant 
while all odd functions change sign. $\partial_+\phi$ will have its parity
conjugate as $-\partial_-\phi$ and consequently a general choice for $L_\pm$
will be
$$L_\pm(x,\lambda)=(T^1(\pm\partial_\pm\phi)+T^2\lambda^{\pm 1} E(\phi)
+T^3\lambda^{\pm 1}(\pm O(\phi))$$
Using Eq.(\ref{11}) we thus obtain
\begin{equation}
L(x,\lambda)=\beta\dot\phi t^1+\frac{m}{2}(\lambda+{\displaystyle\frac{1}{\lambda}})O(\phi)t^2
+\frac{m}{2}(\lambda-{\displaystyle\frac{1}{\lambda}})E(\phi)t^3
\end{equation}
showing the occurrence of canonically conjugate field and hence the possibility
of nontrivial braiding. To get the right asymptotic behaviour we have to
impose the condition $O(\phi)\rightarrow
0$ and $E(\phi)\rightarrow 1$ as $x\rightarrow \pm\infty$. 
To know more about the even and odd 
functions we have to invoke the closure of OPE algebra.  The singularity 
associated with the
OPE of $\dot\phi$ with $O(\phi)$ would involve the derivative $O'(\phi)$ which
is an even function and vice versa. The closure then imposes the condition
\begin{equation}
O'(\phi)\approx E(\phi) {\rm ~~and~~} E'(\phi)\approx O(\phi)
\end{equation}
This means $O$ must be a sine (or sinh) function and $E$ must be a cosine (or 
cosh) function. To study it in detail we start with the OPE
algebra of real scalar field
$$T(\phi(x)\dot\phi(y))=\frac{1}{2\pi(x-y-i0^+)} + ~~\cdots$$
consistent with the equal time commutation algebra
$$[\phi(x^1,x^0),\dot\phi(y^1,x^0)]=i\delta(x^1-y^1).$$
This tells us that 
$$T(O(\phi(x))\dot\phi(y))=\frac{O'(\phi(x))}{2\pi(x-y-i0^+)} + ~~\cdots$$
$$T(E(\phi(x))\dot\phi(y))=\frac{E'(\phi(x))}{2\pi(x-y-i0^+)} + ~~\cdots$$
If we write $O'(\phi(x))=\beta_1E(\phi(x))$ and $E'(\phi(x))=\beta_2O(\phi(x))$
we get, after appropriate redefinition of the generators, two possible form
of $L$, namely,
\begin{equation}
L(x,\lambda)=\beta\dot\phi t^2+\frac{m}{2}(\lambda+{\displaystyle\frac{1}{\lambda}})\sin(\beta\phi)t^1
+\frac{m}{2}(\lambda-{\displaystyle\frac{1}{\lambda}})\cos(\beta\phi)t^3
\label{sglax}
\end{equation}
\begin{equation}
L'(x,\lambda)=\beta\dot\phi t^2+\frac{m}{2}(\lambda+{\displaystyle\frac{1}{\lambda}})\sinh(\beta\phi)t^1
+\frac{m}{2}(\lambda-{\displaystyle\frac{1}{\lambda}})\cosh(\beta\phi)t^3
\end{equation}
Using Eq.(\ref{15}) and exploiting the linear independence of the local fields,
we get the master equations that the braiding matrices ${\cal R}$ for the 
scattering transforms of these two cases
must satisfy if they exist.
\begin{equation}
\begin{array}{l}

[{\cal R},(t^2\otimes 1)+(1\otimes t^2)]=0\\

[{\cal R},(\lambda-{\displaystyle\frac{1}{\lambda}})(t^3\otimes 1)+(\mu-{\displaystyle\frac{1}{\mu}})(1
\otimes t^3)]+{\displaystyle\frac{-\beta^2}{2}}\left\{{\cal R},(\lambda+{\displaystyle\frac{1}{\lambda}})
(t^1\otimes t^2) -(\mu+{\displaystyle\frac{1}{\mu}})(t^2\otimes t^1)\right\}=0\\

[{\cal R},(\lambda+{\displaystyle\frac{1}{\lambda}})(t^1\otimes 1)+(\mu+{\displaystyle\frac{1}{\mu}})(1
\otimes t^1)]+{\displaystyle\frac{\pm \beta^2}{2}}\left\{{\cal R},(\lambda-{\displaystyle\frac{1}{\lambda}})
(t^3\otimes t^2) -(\mu-{\displaystyle\frac{1}{\mu}})(t^2\otimes t^3)\right\}=0
\end{array}\label{sgmaster}
\end{equation}
The $\pm$ in the last equation are for the trigonometric and hyperbolic
cases respectively. We will now invoke the reality condition of the Lax 
operator (a consequence of time reversal invariance) which translates in this
case to the hermiticity of the generators $t^i$. This is guaranteed if we
take the Lie algebra to be compact. Braiding non-triviality also requires that
it should be semi-simple. The only situation with three non-commuting
generators is the $su(2)$ algebra. 

To check the consistency of these equations, we first
notice that for $\beta^2=0$ one has a classical situation, i.e. ${\cal R}=1$.
If the classical limit is smooth, then we can have a perturbative expansion
of ${\cal R}=1+ i\beta^2{\cal R}_1 + \cdots $ and then ${\cal R}_1$ would
satisfy the same relations as above except that the anticommutators in the
right hand side would be replaced  by $-\beta^2[(\lambda+1/\lambda)
(t^1\otimes t^2) -(\mu+1/\mu)(t^2\otimes t^1)]$ and
$\pm \beta^2[(\lambda-1/\lambda)(t^3\otimes t^2)-(\mu-1/\mu)(t^2\otimes t^3)]$.
The most general form of ${\cal R}_1$ consistent with the vanishing commutator
(the first of Eq.(\ref{sgmaster}) is
$${\cal R}_1=a(t^2\otimes t^2)+b(t^3\otimes t^3+t^1\otimes t^1)$$
To determine the unknown coefficients $a$ and $b$ one just substitutes this
in the next equation. The linear independence of $t^1\otimes t^2$ and
$t^2\otimes t^1$ would give rise to two linear equations from which $a$ and
$b$ can be obtained uniquely. This can surely be not consistent with both
forms of the third equation (one for trigonometric and the other for the
hyperbolic)! A few steps of simple algebra shows that only the trigonometric
case is consistent.

If one expands the trace of the scattering transform independently
around
$\lambda=0$ and $\lambda^{-1}=0$ (this is definitely possible in the
classical limit where one can use the usual analytic continuation \cite{AKNS}) 
the co-efficient of $\lambda$ and 
$\lambda^{-1}$ added together would transform like the time component of
momentum vector and hence it can be identified with the Hamiltonian. It 
exactly coincides with the Hamiltonian of a sine-Gordon field.

One thus
concludes that for a single pseudo-scalar field (and its canonical conjugate)
the only consistent quantum field theory permitting a nontrivial braiding of
the scattering transforms is the one whose Lax operator is of the form of
Eq.(\ref{sglax} ). This is our first result. 

The exact solution of Eq.(\ref{sgmaster}) will be representation dependent as 
anti-commutation relation among the generators or their direct products are
not governed by Lie algebra alone. However the exact correspondence of
the degrees of freedom between the original theory (local fields $\phi$ and
$\dot\phi$) and the transformed theory (the off-diagonal elements of the
scattering transform) can be invoked only when one considers a $2\times2$
representation of ${\cal T}$ since there will then be only two off-diagonal 
elements.  In this representation (fundamental $2 \times 2$ 
representation) the master equations can be solved immediately since
different Pauli matrices anti-commute and the solution is
\begin{equation}
\begin{array}{lcl}
{\cal R}(\lambda,\mu)&=&(1\otimes 1)
-i\beta^2\left[ 
{\displaystyle{{\frac{\lambda}{\mu}(1-i\frac{\beta^2}{4})
+\frac{\mu}{\lambda}(1+i\frac{\beta^2}{4})}
\over{\frac{\lambda}{\mu}(1-i\frac{\beta^2}{4})
-\frac{\mu}{\lambda}(1+i\frac{\beta^2}{4})}} (t^2\otimes t^2)}\right.\\
&&\\
&&+\left.{\displaystyle{{2\over{\frac{\lambda}{\mu}(1-i\frac{\beta^2}{4})
-\frac{\mu}{\lambda}(1+i\frac{\beta^2}{4})}}}\left\{(t^3\otimes t^3)+
(t^1\otimes t^1)\right\}}\right]\\
\end{array}\label{Rsg}
\end{equation}
This braiding matrix for the sine-Gordon model has been known for a long
time \cite{Faddeev2, gbsg2} and we just demonstrated here how naturally 
it occurs in relativistic
field theories of pseudo-scalars. Extension to multi-component pseudo-scalars
can be performed by associating sine, cosine functions and the time 
derivatives of each of the pseudo-scalar with the $su(2)$ sub-algebras of the root
systems of an appropriate large Lie algebra. However consistent solutions
are known to exist only when the roots are either simple or the one
associated with the lowest height \cite{Olive}. The Hamiltonian correspond 
to the Toda
model which is basically of sine-Gordon type.

\vspace{24pt}
{\em `Spin' 1/2 Dirac Fields:}

To start with we will assume that all the fields commute rather than 
anti-commute for space like separation. This will ensure that without the
singularities the transposition of the scattering transforms are trivial.
Even if there are anticommuting fields, in one spatial
dimension where a natural ordering exists with the value of the coordinate,
a simple Jordan Wigner transformation
\begin{equation}
\psi\rightarrow \left[e^
{\displaystyle{\pm i\pi\int^{x^1}_{-\infty}j_0(\xi)d\xi}}\right]\psi
\end{equation}
switches a commutator to anti-commutator and vice versa among two complex
fields with space like separation.

The relevant OPE algebras are
\begin{equation}
\psi_i(x^1,x^0)\psi_j^\dagger(y^1,x^0)
={{(-1)^{j+1}\delta_{ij}}\over{2\pi i(x^1-y^1-i(-1)^j0^+)}}+\ 
{\rm less\ singular\ terms}
\label{opeDirac}
\end{equation}
where $i,j$ are the Dirac spinor indices. This is consistent with the 
irreducible LT of the Dirac spinor components given in Eq.(\ref{spinor}) and the
equal time commutation algebra
$$
[\psi_i(x^1,x^0),\psi_j^\dagger(y^1,x^0)]=\delta_{ij}\delta(x^1
-y^1)
$$
A Lorentz covariant (including parity and time reversal) choice of
$L(x,\lambda)$ could be
\begin{equation}
L(x,\lambda)=\left[\frac{m}{2}\left(\lambda
-{\displaystyle\frac{1}{\lambda}}\right)+aj_1\right]t^3
+b\left(\sqrt{\lambda}\psi_1-\sqrt{\displaystyle\frac{1}{\lambda}}\psi_2\right)
t_+
+b^*\left(\sqrt{\lambda}\psi^\dagger_1-\sqrt{{\displaystyle\frac{1}{\lambda}}}
\psi^\dagger_2\right)t_-
\label{Dirac}
\end{equation}
where
$j_1=:\psi^\dagger_1\psi_1:-:\psi^\dagger_2\psi_2:$ is the space component
of the current vector. The two derived OPE relevant for the subsequent study
are
\begin{equation}
T(j_1(x)\psi_k(y))=(-1)^k{{\psi_k(y)}\over{2\pi i(x-y-i0^+)}}+\cdots
\label{jpsiope}
\end{equation}
It should be noted that the OPE of $j_1(x)$ with $j_1(y)$ does not produce any 
singularity in the equal time limit (even though $j_0(x)$ with $j_1(y)$
would have a Schwinger term about which we will discuss later).
While constructing $L$ We could have included higher tensors but that would
not be consistent with the closure property. It should further be noted that 
we have included
only those fields for which the OPE singularities are no more than simple
poles. As a consequence we can proceed, as before, to obtain the set of 
master equations for the braiding matrix for the scattering transform by
comparing the product of infinitesimal strings 
$[{\cal T}_\epsilon(x;\lambda) \otimes 1][1 \otimes {\cal T}_\epsilon(x;\mu)]$ 
with the product in opposite order
after taking into consideration all quantum corrections arising out of
singularities of the OPE given in Eq.(\ref{opeDirac}) and (\ref{jpsiope}). The quantum correction
terms will only have identity, $\psi_{1,2}$ and $\psi^\dagger_{1,2}$ field
operators. The three independent master equations (others essentially
describing the complex conjugation and some discrete symmetry nature of the
braiding matrix) are
\begin{equation}
\begin{array}{l}
\left[{\cal R},(t^3\otimes 1)+(1\otimes t^3)\right]=0\\
\begin{array}{ll}
\left[{\cal R},\left(\lambda-{\displaystyle\frac{1}{\lambda}}\right)(t^3\otimes 1)\right.&
\left. +\left(\mu-{\displaystyle\frac{1}{\mu}}\right)(1
\otimes t^3)\right]\\
&+i|b|^2\left({\displaystyle \sqrt{\lambda\mu}}
+{\displaystyle\frac{1}{\sqrt{\lambda\mu}}}\right)
\left\{{\cal R},(t_-\otimes t_+) -(t_+\otimes t_-)\right\}=0\\
\end{array}\\

\left[{\cal R},\sqrt{\lambda}(t_-\otimes 1)+\sqrt{\mu}(1
\otimes t_-)\right]-{\displaystyle\frac{a}{2}}\left\{{\cal R},\sqrt{\mu}
(t^3\otimes t_-) -\sqrt{\lambda}(t_-\otimes t^3)\right\}=0\\
\end{array}\label{fmaster}
\end{equation}
With the obvious form (consistent with the first equation of Eq.(\ref{fmaster})
$${\cal R}(\lambda,\mu)= 1\otimes 1 + X(\lambda,\mu)t_3\otimes t_3
+ Y(\lambda,\mu)\left[t_+\otimes t_- +t_-\otimes t_+\right]$$
we can solve for the co-efficients $X(\lambda,\mu)$ and $Y(\lambda,\mu)$
from the remaining equations of Eq.(\ref{fmaster}). It turns out, by 
using the linear independence of different $t_i\otimes t_j$, that there
are one too many relations and the consistency can be restored provided
\begin{equation}
|b|^2 =m\left({a\over{1+a^2/16}}\right)
\end{equation}
The exact expression for the braiding matrix turns out to be
\begin{equation}
\begin{array}{ll}
{\cal R}(\lambda,\mu)= 
1\otimes 1 &-ia\left[{\displaystyle
{{\sqrt{\lambda\over\mu}\left(1-{{ia}\over 4}\right)
+\sqrt{\mu\over\lambda}\left(1+{{ia}\over 4}\right)}
\over
{\sqrt{\lambda\over\mu}\left(1-{{ia}\over 4}\right)
-\sqrt{\mu\over\lambda}\left(1+{{ia}\over 4}\right)}}}
\left(t_3\otimes t_3\right)\right.\\
&\left.+
{\displaystyle
{2\over
{\sqrt{\lambda\over\mu}\left(1-{{ia}\over 4}\right)
-\sqrt{\mu\over\lambda}\left(1+{{ia}\over 4}\right)}}
}\left\{\left(t_1\otimes t_1\right) +\left(t_2\otimes t_2\right)\right\}\right]
\end{array}\label{Rmt}
\end{equation}

The existence of the braiding matrix implies the existence of infinite number
of operators in involution. They are the co-efficients of the expansion around
$\lambda=0$ and $\infty$ around which the diagonal elements of ${\cal T}(\lambda
)$ is analytic. The linear combination of the co-efficient of $\lambda$ and
$\lambda_1$ would transform, under LT, like the Hamiltonian. 
A classical evaluation of
this combination yields the Hamiltonian of the massive Thirring model. One
might think at this stage that the model is soluble since the form of the
braiding matrix is exactly known \cite{TB}. But there is a serious problem. With
$2\times 2$ matrix representation of ${\cal T}$ one would land up with only
two scattering transform variables whereas the independent local fields are
four in number ($\psi_1$,$\psi_1$ and their complex conjugates). We will 
address this problem in the next section.

\vspace{24pt}
\noindent{\it Chiral Extension}:

If one relaxes the parity invariance constraint of the scattering transform
and replaces it by parity 
covariance, one gets the interesting possibility of including
true scalar fields in this framework. In this case one will have 
two such transforms, one the parity image of the other, each being
invariant under proper Lorentz transformations. One can now include
both $\partial_x\phi$ and $\dot\phi$ as vector valued fields in $L$. 
For a true scalar field $\phi$, $\partial_x\phi$ is the space component 
of a true vector while $\dot\phi$ is the space component of an axial
vector $\epsilon^{\mu\nu}\partial_\nu\phi$. For the pseudoscalar field
it would just be the opposite. The simplest form of $L$ consistent with
the closure of OPE is of the form
\begin{equation}
L_+(\xi,\lambda)=m\lambda t_3 +(\alpha\partial_x\phi+\beta\dot\phi)t_2
\label{chiral1}
\end{equation}
and its parity counterpart
\begin{equation}
L_-(\xi,\lambda)=\frac{m}{\lambda}t_3 +(-\alpha\partial_x\phi+
\beta\dot\phi)t_2
\label{chiral2}
\end{equation}
The immediate difficulty one encounters here is that the OPE of
$\dot\phi$ and $\partial_x\phi$ now contains a second order pole 
(Schwinger term) and the method prescribed earlier to evaluate the 
quantum correction for the direct product of two strings would not 
work. In literature this is called non-ultralocality. A simple way to get 
around this problem is to identify the ${\cal T}$'s  as  gauge transformation of
another set of $\tilde{\cal T}$'s  whose OPE's will not involve such Schwinger 
terms. Such a possibility was invoked for discrete systems some time back 
\cite{Kundu}.

For example the ${\cal T}(x,y,\lambda)$ associated with Eq.(\ref{chiral1}) can 
be written as
\begin{equation}
{\cal T}_+(x,y,\lambda)=e^{i\alpha\phi(x)t_2}\tilde{\cal T}_+(x,y.\lambda)e^{-i
\alpha\phi(y)t_2}
\end{equation}
with
\begin{equation}
\begin{array}{l}
\tilde{\cal T}_+(x,y.\lambda)={\cal P}e^{i\int^x_y\tilde L_+(\xi,\lambda)d\xi}\\
\tilde L_+(\xi,\lambda)= \beta\dot\phi(\xi) t_2+\lambda\cos(\alpha\phi(\xi))t_3
+\lambda\sin(\alpha\phi(\xi))t_1
\end{array}
\label{gt}
\end{equation}
The exponential factors at the two edges of the string are to be understood
as space-like separated from the fields inside $\tilde{\cal T}$ and hence the whole 
object
is manifestly normal ordered. The local operator $\tilde L_+$ looks very similar
to what we have already discussed earlier and clearly the OPE's among
different $\tilde L_+$'s do not involve singularities above simple poles.  The
quantum correction terms from their OPE's would therefore be similar to
what we have obtained before. There will now be additional correction terms
coming from the OPE of $\dot\phi$ in $\tilde L_+$'s with the exponential terms 
at the edges of the other strings.  To evaluate the contribution to the quantum
corrections coming from the edge terms we follow the convention that in the
OPE of strings the one occurring at the right is infinitesimally shifted down
compared to the one preceding it. Thus we can proceed as before with product
of two infinitesimal strings of length $\epsilon$  
\begin{equation}
\begin{array}{l}
({\cal T}_+(x+\frac{\epsilon}{2},x-\frac{\epsilon}{2} ,\lambda)\otimes 1)(1\otimes 
{\cal T}_+(x+\frac{\epsilon}{2},x-\frac{\epsilon}{2} ,\mu)\\
={\displaystyle :e^{i\alpha\phi(x+\frac{\epsilon}{2})(t_2\otimes 1)}
\left(1\otimes 1+
i\int^{x+\frac{\epsilon}{2}}_{x-\frac{\epsilon}{2}}\big[\tilde L_+(\xi,\lambda)
\otimes 1\big]d\xi\right)e^{-i\alpha\phi(x-\frac{\epsilon}{2})(t_2\otimes 1)}:} 
\\
{\displaystyle :e^{i\alpha\phi(x_-+\frac{\epsilon}{2})(1\otimes t_2)}
\left(1\otimes 1+
i\int^{x_-+\frac{\epsilon}{2}}_{x_--\frac{\epsilon}{2}}\big[1\otimes 
\tilde L_+(\eta,\mu)\big]d\eta\right)
e^{-i\alpha\phi(x_--\frac{\epsilon}{2})(1\otimes t_2)}:}\\
\end{array}
\end{equation}
and use the OPE
$${\displaystyle
e^{iK\phi(\xi)}\dot\phi(\eta)
={{e^{iK\phi(\xi)}}\over{\xi -\eta -i\pi 0^+}}
+:e^{iK\phi(\xi)}\dot\phi(\eta):
}
$$
Because of the shifting convention the edges $x+\frac{\epsilon}{2}$
and $x_--\frac{\epsilon}{2}$ are always space-like separated  from
all the fields occurring within the two strings and hence will not
have any contribution to any quantum correction. Therefore the only
edge contributions come from (a) the product of 
$exp(-i\alpha\phi(x-\frac{\epsilon}{2})(t_2\otimes 1))$ with
$\beta\dot\phi(\eta)(1\otimes t_2)$ and (b) the product of
$\beta\dot\phi(\xi)(t_2\otimes 1)$  with
$exp(i\alpha\phi(x_-+\frac{\epsilon}{2})(1\otimes t_2))$.  

Apart from these edge contributions there are the usual quantum corrections
coming from the OPE's of $\dot\phi$'s with $\sin(\alpha\phi)$'s and
$\cos(\alpha\phi)$'s as discussed earlier. The final result to order $\epsilon$
is
\begin{equation}
\begin{array}{l}
\left({\cal T}_+(x+\frac{\epsilon}{2},x-\frac{\epsilon}{2} ,\lambda)\otimes 1\right)
\left(1\otimes {\cal T}_+(x+\frac{\epsilon}{2},x-\frac{\epsilon}{2} ,\mu)\right)\\
={\displaystyle 
:e^{i\alpha\phi(x+\frac{\epsilon}{2})[(t_2\otimes 1)+(1\otimes t_2)]}
\left[
\left(1\otimes 1 + \frac{i\alpha\beta}{2}(t_2\otimes t_2)\right)^2\right.} \\ 
{\displaystyle +i\int^{x+\frac{\epsilon}{2}}_{x-\frac{\epsilon}{2}}d\xi
\big[\tilde L_+(\xi,\lambda)\otimes 1\big]
\left(1\otimes 1 + \frac{i\alpha\beta}{2}(t_2\otimes t_2)\right)}\\
{\displaystyle +i\left(1\otimes 1 + \frac{i\alpha\beta}{2}(t_2\otimes t_2)\right)
\int^{x+\frac{\epsilon}{2}}_{x-\frac{\epsilon}{2}}d\xi\big[1\otimes\tilde L_+
(\xi,\mu)\big]}\\
{\displaystyle
-\frac{i\alpha\beta}{2}\int^{x+\frac{\epsilon}{2}}_{x-\frac{\epsilon}{2}}d\xi
\big[\cos\alpha\phi(\xi)\big(\lambda(t_1\otimes t_2)-\mu(t_2\otimes t_1)\big)
\big]}\\
{\displaystyle
\left. 
+\frac{i\alpha\beta}{2}\int^{x+\frac{\epsilon}{2}}_{x-\frac{\epsilon}{2}}d\xi
\big[\sin\alpha\phi(\xi)\big(\lambda(t_3\otimes t_2)-\mu(t_2\otimes t_3)\big)
\big]
\right]
e^{-i\alpha\phi(x-\frac{\epsilon}{2})[(t_2\otimes 1)+(1\otimes t_2)]}:
}
\end{array}
\label{lambdamu}
\end{equation}
The product in reverse order can be evaluated in a similar fashion to order
$\epsilon$ (again keeping in mind the convention of shifting the edges).
\begin{equation}
\begin{array}{l}
\left(1\otimes {\cal T}_+(x+\frac{\epsilon}{2},x-\frac{\epsilon}{2} ,\mu)\right)
\left({\cal T}_+(x+\frac{\epsilon}{2},x-\frac{\epsilon}{2} ,\lambda)\otimes 1\right)\\
={\displaystyle 
:e^{i\alpha\phi(x+\frac{\epsilon}{2})[(t_2\otimes 1)+(1\otimes t_2)]}
\left[
\left(1\otimes 1 + \frac{i\alpha\beta}{2}(t_2\otimes t_2)\right)^2
\right.} \\ 
{\displaystyle
+i\left(1\otimes 1 + \frac{i\alpha\beta}{2}(t_2\otimes t_2)\right)
 \int^{x+\frac{\epsilon}{2}}_{x-\frac{\epsilon}{2}}d\xi
\big[\tilde L_+(\xi,\lambda)\otimes 1\big]}\\
{\displaystyle 
+
i
\int^{x+\frac{\epsilon}{2}}_{x-\frac{\epsilon}{2}}d\xi\big[1\otimes\tilde L_+
(\xi,\mu)\big]
\left(1\otimes 1 + \frac{i\alpha\beta}{2}(t_2\otimes t_2)\right)
}\\
{\displaystyle
+\frac{i\alpha\beta}{2}\int^{x+\frac{\epsilon}{2}}_{x-\frac{\epsilon}{2}}d\xi
\big[\cos\alpha\phi(\xi)\big(\lambda(t_1\otimes t_2)-\mu(t_2\otimes t_1)\big)
\big]}\\
{\displaystyle
\left. 
-\frac{i\alpha\beta}{2}\int^{x+\frac{\epsilon}{2}}_{x-\frac{\epsilon}{2}}d\xi
\big[\sin\alpha\phi(\xi)\big(\lambda(t_3\otimes t_2)-\mu(t_2\otimes t_3)\big)
\big]
\right]
e^{-i\alpha\phi(x-\frac{\epsilon}{2})[(t_2\otimes 1)+(1\otimes t_2)]}:
}
\end{array}
\label{mulambda}
\end{equation}
To find the braiding relation, if any, one has to solve for a ${\cal R}$ matrix that
would connect Eq.(\ref{lambdamu}) with Eq.(\ref{mulambda}).  This would,
in turn, by virtue of the linear independence of the fields,
lead to a set of algebraic master equations for R.

As before, comparison of terms proportional to $\dot\phi$ gives
\begin{equation}
\left[{\cal R}(\lambda,\mu), (t_2\otimes 1)+(1\otimes t_2)\right]=0
\end{equation}
and comparison of terms proportional to $\cos(\alpha\phi)$ gives
\begin{equation}
\left[{\cal R}(\lambda,\mu), \lambda(t_3\otimes 1)-\mu(1\otimes t_3)\right]
=\frac{\alpha\beta}{4}\left\{{\cal R},\lambda(t_1\otimes t_2)-\mu(t_2\otimes t_1)
\right\}
\end{equation}
Comparison of terms proportional to $\sin(\alpha\phi)$ gives no new relation
(only reflecting the automorphism symmetry of the $t$'s).

The master equations are very similar to the one we obtained for the 
sine-Gordon theory (Eq.(\ref{sgmaster}. The coupling constant is $\alpha\beta/4$ instead of
$\beta^2/2$ and the spectral parameters are directly $\lambda$ and $\mu$
instead of momentum and energy. The resultant $R(\lambda,\mu)$ however
has identical form
\begin{equation}
\begin{array}{l}
{\cal R}(\lambda,\mu)=1\otimes 1 + A(\lambda,\mu)t_2\otimes t_2
+B(\lambda,\mu)(t_3\otimes t_3+t_1\otimes t_1)\\
{\displaystyle 
A(\lambda,\mu)=-\frac{i\alpha\beta}{2}{{
\frac{\lambda}{\mu}\left(1-\frac{i\alpha\beta}{8}\right)
+\frac{\mu}{\lambda}\left(1+\frac{i\alpha\beta}{8}\right)}\over
{\frac{\lambda}{\mu}\left(1-\frac{i\alpha\beta}{8}\right)
-\frac{\mu}{\lambda}\left(1+\frac{i\alpha\beta}{8}\right)   
}}
}\\
{\displaystyle 
B(\lambda,\mu)=-\frac{i\alpha\beta}{2}{
{2}\over
{\frac{\lambda}{\mu}\left(1-\frac{i\alpha\beta}{8}\right) 
-\frac{\mu}{\lambda}\left(1+\frac{i\alpha\beta}{8}\right)   
}}
}
\end{array}
\label{Rplus}
\end{equation}
That the scattering transform corresponding to Eq.(\ref{chiral1}) 
would lead to a classical integrable system similar
to the sine-Gordon theory was well-known \cite{AKNS} and therefore
it is not surprising that the braiding matrix for quantum theory of them 
would be of the same form.
The only difference is that the parity conservation would
make it necessary to use the scattering transforms corresponding to 
both Eq.(\ref{chiral1}) and Eq.(\ref{chiral2}) to describe one
single system. It is therefore a chiral theory having twice the degrees of
freedom compared to the ordinary sine-Gordon theory.  That it is really
a chiral theory can be checked by the following three steps.

(a) The scattering transforms $T_+$ corresponding to Eq.(\ref{chiral1}) have 
already been shown to satisfy the braiding relation
\begin{equation}
{\cal R}_+(\lambda,\mu)\left[{\cal T}_+(x,y;\lambda)\otimes 1\right]
\left[1\otimes {\cal T}_+(x,y;\mu)\right]
=
\left[1\otimes {\cal T}_+(x,y;\mu)\right]\left[{\cal T}_+(x,y;\lambda)\otimes 1\right]
{\cal R}_+(\lambda,\mu)
\end{equation}
with ${\cal R}_+(\lambda,\mu)$ given by Eq.(\ref{Rplus}).

(b) In an identical way one can obtain the braiding relations among the
scattering transforms ${\cal T}_-$ corresponding to Eq.(\ref{chiral2}) by noting
that this too is a gauge transform of an ultra-local theory
\begin{equation}
{\cal T}_-(x,y,\lambda)=e^{-i\alpha\phi(x)t_2}\tilde {\cal T}_-(x,y.\lambda)e^{i\alpha\phi(y)t_2}
\end{equation}
with
\begin{equation}
\begin{array}{l}
\tilde {\cal T}_-(x,y.\lambda)={\cal P}e^{i\int^x_y\tilde L_-(\xi,\lambda)d\xi}\\
\tilde L_-(\xi,\lambda)= \beta\dot\phi(\xi) t_2+\lambda\cos(\alpha\phi(\xi))t_3
-\lambda\sin(\alpha\phi(\xi))t_1
\end{array}
\label{gt2}
\end{equation}

This would lead to
\begin{equation}
{\cal R}_-(\lambda,\mu)\left[{\cal T}_-(x,y;\lambda)\otimes 1\right]
\left[1\otimes {\cal T}_-(x,y;\mu)\right]
=
\left[1\otimes {\cal T}_-(x,y;\mu)\right]\left[{\cal T}_-(x,y;\lambda)\otimes 1\right]
{\cal R}_-(\lambda,\mu)
\end{equation}
with ${\cal R}_-(\lambda,\mu)$ having the similar form as of ${\cal R}_+(\lambda,\mu)$
except $\alpha\beta$ is to be replaced by $-\alpha\beta$ and $\lambda$
interchanged with $\mu$.

(c) The mixed braiding among ${\cal T}_+$ and ${\cal T}_-$ turn out to be trivial in the
sense that the braiding matrix does not depend on the spectral parameters.
This is an unexpected result and needs a little elaboration. Notice the 
differences in signs before the $\sin(\alpha\phi)$ in the Eq.(\ref{gt})
and Eq.(\ref{gt2}). It is this feature that is reflected in the quantum
correction in the mixed product in a form slightly different from what we
have got in Eq.(\ref{lambdamu}) and in Eq.(\ref{mulambda}). If one uses
the relations (true only in the fundamental representation of the $t$'s)
$$(t_2\otimes t_3)=\frac{-i}{2}(1\otimes t_1)(t_2\otimes t_2),$$
$$(t_2\otimes t_1)=\frac{i}{2}(1\otimes t_3)(t_2\otimes t_2),$$ 
and the fact that $(t_2\otimes t_2)$ anticommutes with $(1\otimes t_{1,3})$
as well as $(t_{1,3}\otimes 1)$, we will get the master equations for
the ${\cal R}_{\rm mixed}$ by comparing separately the co-efficients of the fields
as
\begin{equation}
\begin{array}{l}
\left[{\cal R}_{\rm mixed},(t_2\otimes 1)\pm(1\otimes t_2)\right]=0,\\
\left[i{\cal R}_{\rm mixed},\lambda(t_3\otimes 1) +\mu(1\otimes t_3)\right]\\
+\left[\frac{\alpha\beta}{2}(t_2\otimes t_2)R_{\rm mixed},
\lambda(t_3\otimes 1) +\mu(1\otimes t_3)\right]=0,\\
\left[i{\cal R}_{\rm mixed},\lambda(t_1\otimes 1) -\mu(1\otimes t_1)\right]\\
+\left[\frac{\alpha\beta}{2}(t_2\otimes t_2){\cal R}_{\rm mixed},
\lambda(t_1\otimes 1) -\mu(1\otimes t_3)\right]=0
\end{array}
\end{equation}
These equations tell that $(1-\frac{i\alpha\beta}{2}(t_2\otimes t_2))
R_{\rm mixed}$ must be proportional to identity. Consequently, since
overall normalisation of $R$ is immaterial, we get
\begin{equation}
{\cal R}_{\rm mixed}=1+\frac{i\alpha\beta}{2}(t_2\otimes t_2)
\end{equation}

The triviality of ${\cal R}_{\rm mixed}$ means that the all the elements of the
scattering transforms of one chirality essentially commute with all the
elements of the other chirality. This is what is really expected out of
a two-component chiral theory. The fact that the action variables (the
diagonal elements of ${\cal T}$ - the constants of motion) of one chirality commutes
with the angle variables (the non-diagonal elements - the creation and
destruction operators) of the other chirality implies that the chiral states
do not mix during time evolution. In spite of a mass scale present in the
theory the model still exhibits a characteristic of a massless theory,
namely, the conservation of chirality ($\gamma^5$ invariance).

A similar chiral extension is also possible for a Dirac Theory of
Eq.(\ref{Dirac}) where $gj_1$ can be replaced by $gj_1\pm\alpha j_0$.
A very general expression for a chiral $L$ could be
\begin{equation}
L(x,\lambda)=\left(k_1 +\gamma^5\beta k_0
+aj_1 + \gamma^5\alpha j_0\right)t_3
+\beta'\left(\sqrt{\lambda}\psi_1-\beta''\gamma^5\sqrt{\displaystyle\frac{1}{\lambda}}\psi_2\right)
t_+ + {\rm h. c.}
\label{chiralDirac}
\end{equation}
Once again one can gauge relate such a theory with the integrable system which 
we have already discussed in the parity invariant Dirac case. For a consistent
solution for the braiding matrix it turns out that the new parameters
$\beta,\ \beta',\ \beta''$ and $\alpha$ are related. Now one will 
have a two component chiral description 
for both the creation and destruction parts and their algebra with the 
two-fold sets of constants of motion. If such a problem can be quantised 
consistently one would have a true solution for a Massive Thirring model.
In contrast to the case of scalar chiral fields this time, however, it would 
not describe a decoupled theory. The detailed dynamical studies of such 
two-fold chiral integrable systems would be carried out in a latter article.

\end{document}